\documentclass[12pt]{article}
\usepackage{amsmath,amssymb,graphicx}

\begin{document}

\title{A simple effective method for curvatures estimation on triangular meshes}
\date{Sep, 2005}\maketitle {\center

\author{
            Jyh-Yang Wu\footnote{Department of Mathematics , National Chung Cheng University, Chia-Yi 621, Taiwan.
                                 Email:jywu@math.ccu.edu.tw}
           Sheng-Gwo Chen\footnote{Department of Applied Mathematics, National ChiaYi University, Chia-Yi 600, Taiwan.
                                  Email:csg.chen@msa.hinet.net},
       and Mei-Hsiu Chi \footnote{Department of Mathematics , National Chung Cheng University, Chia-Yi 621, Taiwan.
                                 Email:mhchi@math.ccu.edu.tw}}\\

\begin{abstract}
To definite and compute differential invariants, like curvatures,
for triangular meshes (or polyhedral surfaces) is a key problem in
CAGD and the computer vision. The Gaussian curvature and the mean
curvature are determined by the differential of the Gauss map of
the underlying surface. The Gauss map assigns to each point in the
surface the unit normal vector of the tangent plane to the surface
at this point. We follow the ideas developed in Chen and Wu
\cite{Chen2}(2004) and Wu, Chen and Chi\cite{Wu}(2005) to describe
a new and simple approach to estimate the differential of the
Gauss map and curvatures from the viewpoint of the gradient and
the centroid weights. This will give us a much better estimation
of curvatures than Taubin's algorithm \cite{Taubin} (1995).

\end{abstract}

}

\section{Introduction}

The tensors of curvatures on a regular surface $\Sigma$ in the 3D
Euclidean space are important differential invariants in the
theory of surfaces and its applications. Gaussian curvatures and
mean curvatures are determined by the differential of the Gauss
map on the surface $\Sigma$. Given a basis on the tangent plane of
a point p on the surface $\Sigma$, the differential $dN$ of the
Gauss map $N$ can be realized as a $2\times2$ symmetric matrix
$A$. In fact, the Gaussian curvature and the mean curvature at the
point $p$ can be computed from the determinant and the trace of
the matrix $A$. Since 1990, many methods, like Chen and Schmitt
\cite{Chen1}(1992) and Taubin \cite{Taubin} (1995), to estimate
these curvatures are proposed. However, most of them are devoted
to the investigation of the principal curvatures, but not directly
from the differential of the Gauss map. Usually, the accurate
estimation of curvatures at vertices on a triangular mesh plays as
the first step for many applications such as simplification,
smoothing, subdivision, visualization and registration, etc.

Chen and Schmitt \cite{Chen1} (1992) and Taubin
\cite{Taubin}(1995) employed the circular arcs to approximate the
normal curvatures. Their methods need to estimate the Euler
formula and may cause large errors. In this note, we follow the
methods developed by Chen and Wu \cite{Chen2,Chen3}(2004, 2005)
and Wu, Chen and Chi \cite{Wu}(2005) to estimate the differential
$dN$ of the Gauss map directly and then obtain accurate
estimations of the Gaussian curvature and the mean curvature. This
method follows the line of the differential of the Gauss map more
directly and thus provides us a conceptually simple algorithm to
estimate curvatures on a triangular mesh. Since our method is more
natural, the estimation turns out to be more accurate. Indeed, it
performs much better than many other proposed methods. In section
two we recall the basic theory about the differential of Gauss map
on a regular surface. In section three, we briefly review the
methods of Chen and Schmitt, and Taubin to estimate curvatures on
triangular meshes. In section four, we present our method for
estimating the differential of the Gauss map and curvatures on a
triangular mesh. In section five, we compare the results of our
method with Taubin's method.

\section{The Gauss map and curvatures on regular surfaces}

Consider a parameterization $x: U \rightarrow \Sigma$ of a regular
surface $\Sigma$ at a point $p$, where $U$ is an open subset of
the 2D Euclidean space $\mathbb{R}^2$. We can choose, at each
point $q$ of $x(U)$, an unit normal vector $N(q)$. The map $N:
x(U)\rightarrow S^2$ is the local Gauss map from an open subset of
the regular surface $\Sigma$ to the unit sphere $S^2$ in the 3D
Euclidean space $\mathbb{R}^3$. The Gauss map $N$ is
differentiable and its differential $dN$ of $N$ at $p$ in $\Sigma$
is a linear map from the tangent space into itself.

Given an orthogonal basis $\{ e_1, e_2 \}$ for the tangent space
$T\Sigma_p$, we can find a  $2 \times 2$ matrix $A$ to represent
the Gauss map as follows: set $a_{ij}$ to be the inner product of
the vector $e_i$ with $dN_p(e_j)$. That is,
\begin{equation}
a_{ij} = < e_i , dN_p(e_j)>.
\end{equation}
Thus the matrix $A=(a_{ij})$ represents the linear map $dN_p$.
Given a vector $v = \alpha_1 e_1+ \alpha_2 e_2$ in $T\Sigma_p$,
$dN_p(v)$ is equal to $w = \beta_1e_1 + \beta_2e_2$ where
$\beta_1$ and $\beta_2$ satisfy

\begin{equation}
\begin{pmatrix} \beta_1 \cr \beta_2 \end{pmatrix} =
\begin{pmatrix} a_{11} & a_{12} \cr a_{21} & a_{22} \end{pmatrix}
\begin{pmatrix} \alpha_1 \cr \alpha_2 \end{pmatrix}
\end{equation}
 Under this representation of the linear map $dN_p$, the Gaussian curvature $K$ and
the mean curvature $H$ can be computed by $K=\det(A)$ and
$H=-\frac{\mbox{ trace}(A)}{2}$. Namely,

\begin{equation}
K = a_{11}a_{22} - a_{12}a_{21}
\end{equation}

\begin{equation}
H = -\frac{a_{11}+a_{22}}{2}
\end{equation}

Indeed, since the linear map $dN_p$ is self-adjoint, the matrix
$A$ should be symmetric and diagonalizable. Its eigenvalues are
$-\kappa_1$ and  $-\kappa_2$. The values $\kappa_1$ and $\kappa_2$
are the principal curvatures and their associated eigenvectors
$v_1$ and $v_2$ are called the principal directions. In terms of
the principal curvatures $\kappa_1$ and $\kappa_2$, we have the
Gaussian curvature $K= \kappa_1\kappa_2$ and the mean curvature $H
= \frac{\kappa_1+\kappa_2}{2}$.

Consider a regular curve $c(s)$ with arc length in the regular
surface $\Sigma$ and $c(0)=p$. The number $k_n=<c''(0),N(p)>$ is
called the normal curvature of $c(s)$ at $p$. Meusnier's Theorem
\cite{Do}(1976) implies that all curves at $p$ with the same
tangent vector will have the same normal curvature. This allows us
to speak of the normal curvature along a unit tangent vector at
$p$. The maximum and the minimum normal curvatures are nothing but
the principal curvatures $\kappa_1$ and $\kappa_2$. Moreover, we
have, for any unit vector $v$ in $T \Sigma_p$,

\begin{equation}
 v = v_1 \cos \theta + v_2 \sin \theta
\end{equation}

where $v_1$ and $v_2$ are the principal directions and $\theta$ is
the angle between the vectors $v$ and $v_1$. The normal curvature
$k_n$  along the unit vector $v$ is given by

\begin{equation}\label{euler}
k_n(v) = \kappa_1 \cos^2 \theta + \kappa_2 \sin^2 \theta
\end{equation}
This is known as the Euler formula.

\section{The normal curvature approach: Chen and Schmitt's method and Taubin's method}

In this section we shall review the methods given by Chen and
Schmitt \cite{Chen1}(1992) and Taubin \cite{Taubin}(1995). First
we introduce some notations. Consider a triangular mesh $S=(V,F)$,
where $V=\{ v_i|1\leq i \leq n_v \}$ is the list of vertices and
$F=\{f_k|1\leq k \leq n_F \}$ is the list of triangles. We assume
$S$ is oriented and consistent. That is, neighboring triangles
have the normals pointing to the same side of the surface. For a
given vertex $v$ in $V$, we say that another vertex $w$ is a
neighbor vertex of $v$ if there is a triangle $f$ in $F$ such that
$v$ and $w$ are both in $f$. We denote by $Ng(v)$ the set of
neighbor vertices $w$ of $v$ in $V$ and $m$ the number of points
in $Ng(v)$. We also denote by $T(v)$ the set of triangles $f$ in
$F$ with $v \in f$. If the triangle $f$ is in $T(v)$, we say that
$f$ is incident to $v$. The area of a triangle $f$ is denoted by
$|f|$.

\begin{figure}[htb]
{\center
\includegraphics[scale=.9]{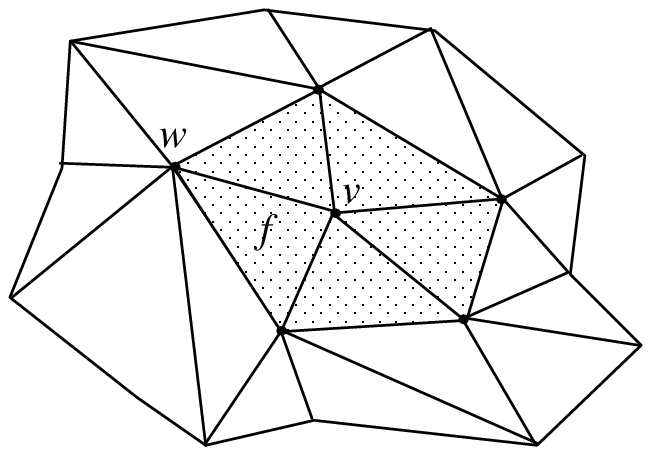}
}
\end{figure}

\subsection{The least square method of Chen and Schmitt}

Chen and Schmitt \cite{Chen1}(1992) provided an algorithm to
estimate the principal curvatures by the Euler formula. Their main
idea is to choose a suitable coordinate system $\{ r_1, r_2 \}$ on
the tangent space. Given a unit vector t in the tangent plane, the
Euler formula (\ref{euler}) will give
\begin{equation}\label{euler2}
k_n(t) = \kappa_1 \cos^2 (\theta + \theta_0) + \kappa_2 \sin^2
(\theta + \theta_0)
\end{equation}
 where $\theta_0$ is the angle between the principal direction
 $e_1$ and $r_1$ and $\theta$ is now the angle between the unit vector $t$ and $r_1$. This
equation (\ref{euler2}) can be rewritten as

\begin{equation}
k_n(t) = C_1 \cos^2 \theta + C_2 \cos^2 \theta + C_3 \cos^2 \theta
\end{equation}
 for some constant $C_1$, $C_2$ and $C_3$.

Given a vertex $v_i$ in $Ng(v)$, we can obtain an unit vector
$t_i$ in the tangent plane of $S$ at the vertex $v$ by

\begin{equation}\label{tangent}
t_i = \frac{(v_i - v) - <v_i-v,N>N}{||(v_i - v) - <v_i-v,N>N||}
\end{equation}
where $N$ is the normal vector at $v$. The normal curvature
$k_n(t)$ along the unit vector $t_i$ can now be approximated by
\begin{equation}\label{kn}
k_n(t_i) = \frac{2<v_j-v,N>}{||v-v_i||^2}.
\end{equation}
See Chen and Wu \cite{Chen2}(2004) for the discussions. Chen and
Schmitt used the least square method to find the constants $C_1$,
$C_2$ and $C_3$:

\begin{equation}
\min \sum_i \left | (C_1 \cos^2 \theta_i + C_2 \cos \theta_i \sin
\theta_i + C_3 \sin^2 \theta_i) - k_n(t_i) \right |^2
\end{equation}

where $\theta_i$ is the angle between $t_i$ and $r_1$. The
principal curvatures can then be solved from the constants $C_1$,
$C_2$ and $C_3$ via the following relations:

\begin{equation}\left\{
\begin{array}{ll}
\kappa_1 \cos^2 \theta_0 + \kappa _2 \sin^2 \theta_0 & = C_1 \cr
 2(\kappa _1 - \kappa _2) \cos \theta_0 \sin \theta_0 & = C_2 \cr
\kappa _2 \cos^2 \theta_0 + \kappa _1 \sin^2 \theta_0 & = C_3
\end{array}\right.
\end{equation}

Moreover, the principal directions can also be computed from
\begin{equation}\left\{
\begin{array}{ll}
v_1 & = \cos (-\theta_0)r_1 + \sin(-\theta_0)r_2 \cr
 v_2 & = \sin (-\theta_0)r_1 + \cos(-\theta_0)r_2
\end{array}\right.
\end{equation}

\subsection{The integral method of Taubin}

To find the principal curvatures $\kappa_1$ and  $\kappa_2$,
Taubin \cite{Taubin}(1995) considered the following integral of a
symmetric $3 \times 3$ matrix:

\begin{equation}\label{Int1}
B = \frac{1}{2} \int_{-\pi}^{\pi} k_n(v)vv^T d\theta
\end{equation}

where $v=v_1 \cos \theta + v_2 \sin \theta$. Taubin showed that
the matrix $B$ can be decomposed into

\begin{equation}
B = E^T \begin{pmatrix} m_1 & 0 \cr 0 & m_2 \end{pmatrix} E
\end{equation}
where $ E = \begin{pmatrix} v_1 & v_2 \end{pmatrix}$ is the $3
\times 2$ matrix given by the principal directions. Moreover, the
principal curvatures are given by the relations:

\begin{equation}
\left\{
\begin{array}{ll}
\kappa_1 & = 3m_1 - m_2 \cr \kappa_2 & = 3m_2 - m_1
\end{array}\right.
\end{equation}

Taubin approximated the integral (\ref{Int1}) by the finite sum

\begin{equation}\label{Taubin1}
 \tilde{B} = \sum_{i=1}^m \omega_i k_n(t_i) t_it_i^T,
\end{equation}
where the unit tangent vector $t_i$ is given in (\ref{tangent})
and the normal curvature is approximated as in equation
(\ref{kn}). The weight $\omega_i$ is chosen to be proportional to
the sum of the areas of the triangles incident to both $v$ and
$v_i$ with $\sum_{i=1}^m \omega_i = 1$. Taubin decomposed the
matrix $\tilde{B}$ with a suitable transformation and a rotation.
In (Chen and Wu \cite{Chen2}), the centroid weights were used for
Equation (\ref{Taubin1}) and they gave more accurate results.

\subsection{ Related works}

Flynn and Jain \cite{Flynn}(1989) used a suitable sphere passing
through four vertices to estimate curvatures. Meek and Walton
\cite{Meek}(2000) examined several methods and compared them with
the discretization and interpolation method. Gatzke and
Grim\cite{Gatzke}(2003) systematically analyzed the results of
computation of curvatures of surfaces represented by triangular
meshes and recommended the surface fitting methods. See also
Petitjean \cite{Petitjean}(2002) for the surface fitting methods.
Meyer et al.\cite{Meyer}(2003) employed the Gauss-Bonnet theorem
to estimate the Gaussian curvatures and introduced the
Laplace-Beltrami operator to approximate the mean curvature.

\section{ The differential approach: our method}

In this section we shall describe a new, simple and effective
method to approximate the differential of the Gauss map. In order
to simplify the presentation, we shall follow the ideas developed
in Wu, Chen and Chi\cite{Chen3}(2005) about the gradient and the
Laplacian of a function defined on a triangular mesh.

Consider a triangular mesh $S=(V,F)$, where $V=\{v_i | 1 \leq i
\leq n_v \}$ is the list of vertices and $F=\{f_k | 1 \leq i \leq
n_F \}$ is the list of triangles. Let $g$ be a function on $V$.
First, we can extend the function $g$ to a piecewise linear
function, still denoted by $g$, on $S$ as follows. Given a face
$f$ in $F$ with vertices $v_i$, $v_j$ and $v_k$, every point $p$
in $f$ can be written as an unique linear combination of $v_i$,
$v_j$ and $v_k$. That is, $p=\alpha v_i + \beta v_j + \gamma v_k$
with $\alpha, \beta, \gamma \geq 0$ and $\alpha + \beta + \gamma =
1$. Then we define $g(p)$ by
\begin{equation}
g(p) = \alpha g(v_i) + \beta g(v_j) + \gamma g(v_k)
\end{equation}

Thus, the function $g$ is affine on each face $f$ of $S$ and hence
is differentiable on $f$. The gradient  $(\nabla g)_f$ of $g$ on
the face $f$ at the vertex $v_i$ can be computed from
\begin{equation}
(\nabla g)_f (v_i) = a(v_j - v_i) + b(v_k - v_i)
\end{equation}
where the coefficients $a$ and $b$ are determined by the relations
\begin{equation}
\begin{array}{ll}
g(v_j) - g(v_i) &= < (\nabla g)_f (v_i), v_j-v_i>, \cr
g(v_k) - g(v_i) &= < (\nabla g)_f (v_i), v_k-v_i>, \cr
\end{array}
\end{equation}

A direct computation gives
\begin{equation}
\begin{pmatrix} a \cr b \end{pmatrix} =
\begin{pmatrix}
<v_j-v_i,v_j-v_i>, & <v_j-v_i,v_k-v_i> \cr <v_j-v_i,v_k-v_i>, &
<v_k-v_i,v_k-v_i>
\end{pmatrix}^{-1} \begin{pmatrix}
g(v_j) - g(v_i) \cr g(v_k) - g(v_i)
\end{pmatrix}
\end{equation}

To obtain the gradient $\nabla g(v_i)$ of $g$ on $S$ at the vertex
$v_i$, we use the weighted combination method. Namely, we set
\begin{equation}\label{gradient}
\nabla g(v_i) = \sum_{f \in T(v_i)} \omega _f [(\nabla g)_f(v_i)]
\end{equation}

with $\omega _f \geq 0 $ and $\sum_{f \in T(v_i)} \omega _f =1$.
According to Chen and Wu \cite{Chen2,Chen3}(2004, 2005), we shall
use the centroid weight for the gradient formula (\ref{gradient}).
The centroid weights are
\begin{equation}\label{centroid}
\omega_f = \frac{\frac{1}{||G_f - v_i||^2}}{\sum_{f \in
T(v_i)}\frac{1}{||G_f - v_i||^2}}
\end{equation}

 where the centroid $G_k$ of the triangle face $f$ is determined by
 \begin{equation}
G_f = \frac{v_i+v_j+v_k}{3}
 \end{equation}

Next we consider the differential $dN$ of the Gauss map $N$. As in
Chen and Wu\cite{Chen2}(2004), we define the normal vector $N(v)$
at each vertex $v$ in $V$ by the centroid weights:
\begin{equation}
N(v) = \frac{\sum_{f \in T(v)} \omega_f N_f}{\left \| \sum_{f \in
T(v)} \omega_f N_f\right \|}
\end{equation}
where $N_f$ is the unit normal to the triangle face $f$ and the
weight $\omega_f$ is given in (\ref{centroid}). The normal vector
$N(v)$ has three components:
\begin{equation}
N(v) = \begin{pmatrix} n_1(v), n_2(v), n_3(v) \end{pmatrix}^T
\end{equation}
and its components $n_i(v)$ are functions on $V$. Thus, we can
compute their gradients and obtain the differential $dN_v$ of $N$
at the vertex $v$ as
\begin{equation}
dN_v = \begin{pmatrix} \nabla n_1(v), \nabla n_2(v), \nabla n_3(v)
\end{pmatrix}.
\end{equation}

Note that the differential $dN_v$ is a $3 \times 3$ matrix. In
order to obtain the linear map $dN_v$ from the tangent space
$TS_v$ into itself, we choose an orthonormal basis $\{ e_1, e_2
\}$ for the tangent space $T\Sigma _p$. That is, the vectors
$e_1$, $e_2$ and $N(v)$ form an 3-dimensional orthonormal basis.
Then the differential $dN_v$ can be realized by a $2 \times 2$
matrix, still denoted by $dN_v = \begin{pmatrix} a_{ij}
\end{pmatrix}$,
 and the entry $a_{ij}$ is given by
\begin{equation}
a_{ij} = < e_i, dN_v(e_j) >.
\end{equation}
Therefore, we can estimate the Gaussian curvature $K$ and the mean
curvature $H$ by

\begin{equation}
\begin{array}{ll}
K &= a_{11}a_{22} - a_{12}a_{21} \cr H & =
-\frac{a_{11}+a_{22}}{2}
\end{array}
\end{equation}

The principal curvatures and the principal directions can also be
computed from the eigenvalues and eigenvectors of $dN_v$.

\section{Computational Results}
Taubin's method to estimate the tensors product of curvature on
triangular mesh is very useful in CAD. Furthermore, Chen and Wu
\cite{Chen1} (2004) provided a better choice in Taubin's algorithm
by centroid weights. We will compare Taubin's method by centroid
weights and our method. In our tests, we consider the random
polynomial surface,
\begin{equation}\label{random_surface}
 \Sigma = \{ (u,v,f(u,v)) \}
\end{equation}
with
$$ f(u,v ) = \sum_{i=0}^{m} \sum_{j=0}^n c_{ij} u^iv^j $$
 We
compute the Gaussian curvature at the vertex $p = (0,0,f(0,0))$ on
some different random surfaces. The set of neighbors of $p$ is
constructed by
\begin{equation}\label{random_partition}
 \{ ( r_i \cos \theta_i , r_i \sin \theta_i , f(r_i
\cos \theta_i , r_i \sin \theta_i) ) | i \in \{ 1,2, \cdots, n_V
\} \}
\end{equation}

In equations \ref{random_surface} and \ref{random_partition},
$c_{ij}$ is a random number in the interval $[ -5,5]$,  $\{
\theta_0 , \theta_1 , \cdots \theta _{n_V} \}$ is a random
partition of $[0,2\pi]$ such that $  0 < |\theta _{j+1} - \theta_j
| < 1.9 \pi $ for each $j$ and $m, n, r_i, n_V \in \mathbb{R}^{+}$
are some random positive values. And we estimate the error of
Gaussian curvatures by the formula
$$ Err(K) = \frac{|K-K_{v}|}{K}$$
where $K$ is the real Gaussian curvature at vertex $v$ and $K_v$
is the Gaussian curvature at vertex $v$ obtained by Taubin's
method or our method.

\begin{figure}
{\center
\includegraphics[scale=.8]{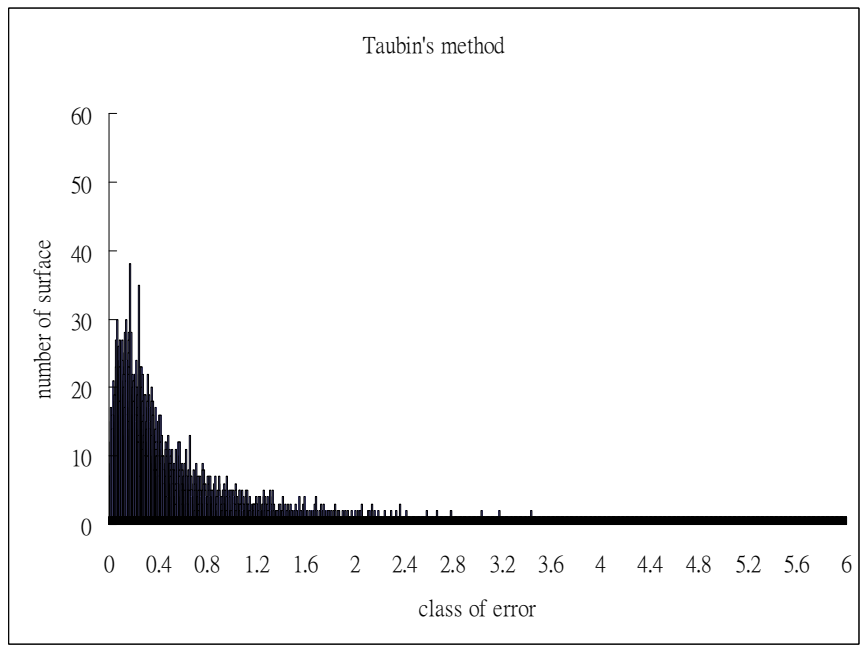}
\caption{The effect of different surface of Taubin's
method}\label{result1} }\end{figure}

\begin{figure}
{\center
\includegraphics[scale=.8]{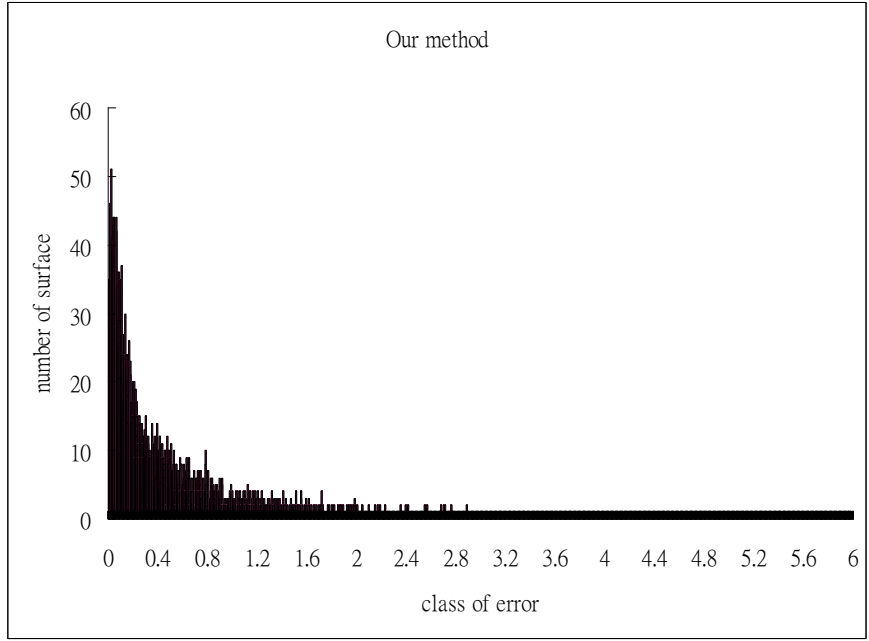}
\caption{The effect of different surface of Our
method}\label{result2} }\end{figure}

\begin{figure}
{\center
\includegraphics[scale=.6]{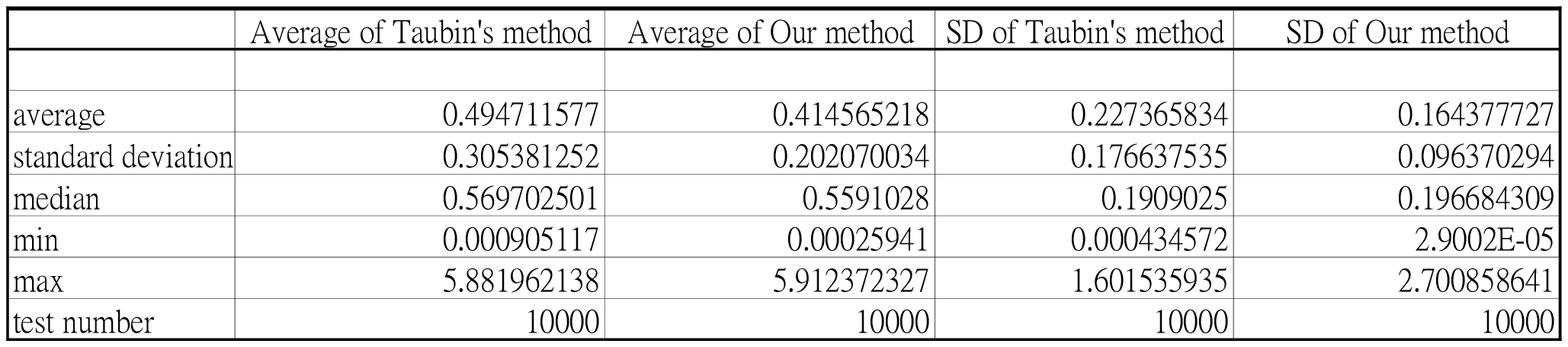}
\caption{The table of the effect of different
surface}\label{result3} }\end{figure}

\begin{figure}
{\center
\includegraphics[scale=.8]{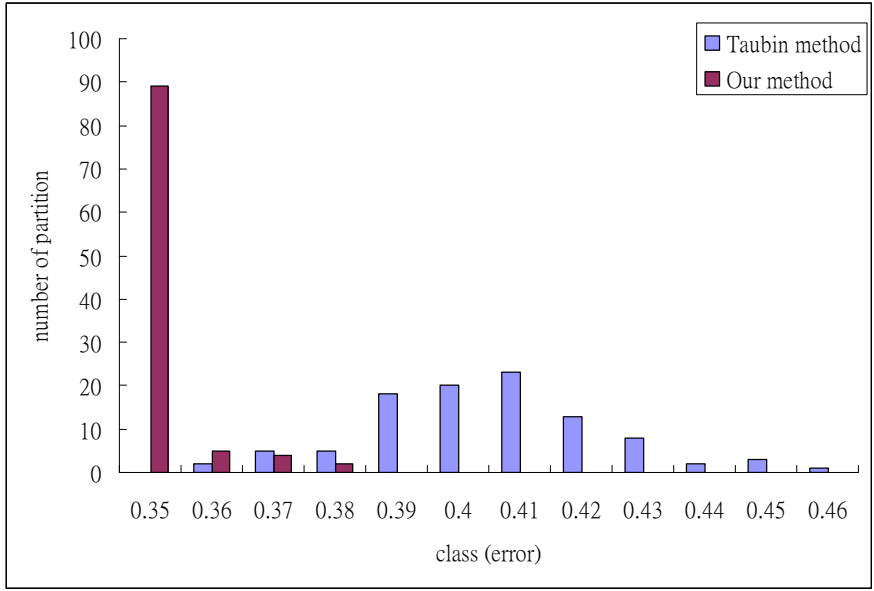}
\caption{The effect of different partition}\label{result4}
}\end{figure}

\begin{figure}
{\center
\includegraphics[scale=.8]{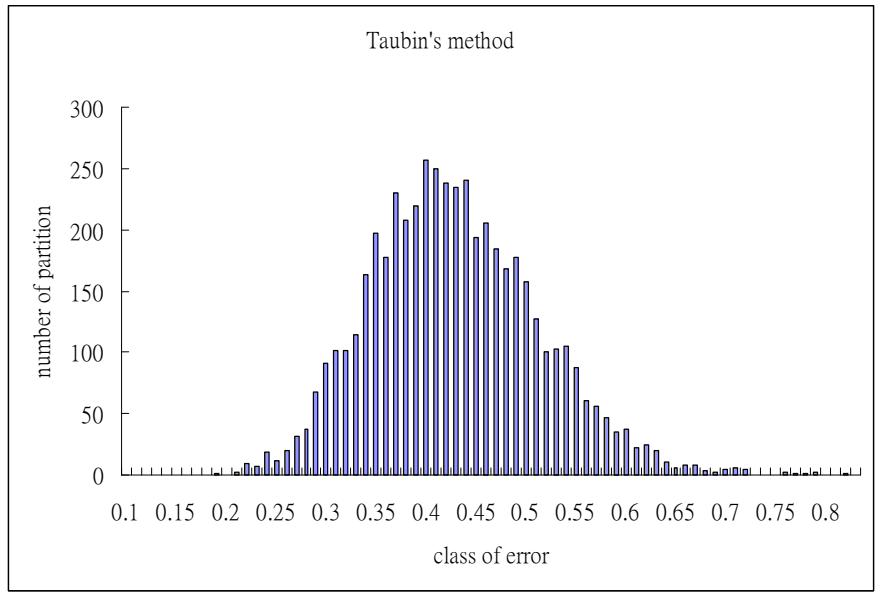}
\caption{The effect of different partition of Taubin's
method}\label{result5} }\end{figure}

\begin{figure}
{\center
\includegraphics[scale=.8]{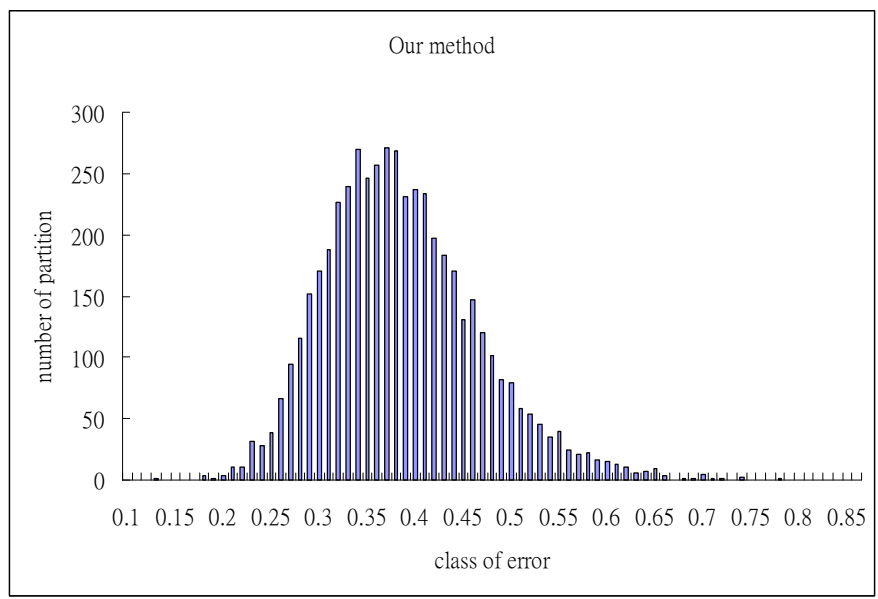}
\caption{The effect of different partition of our
method}\label{result6} }\end{figure}

From figure \ref{result1} to \ref{result3}, we tests 1,000 random
surfaces. For each surface, we compute the error of average of
10,000 different random partitions. From these figures, our method
is better than the Taubin's method. In Figures \ref{result4},
\ref{result5} and \ref{result6}, we test the effect of different
partitions. For each partition, we choose 10,000 differential
random surfaces and estimate the error of average and standard
derivation. Obviously, Our method is more stable than the Taubin's
method.

{\bf Final Remark:} \\
 This method is conceptually simple and more
natural than the normal curvature method of Chen and Schmitt and
Taubin. In the next section, we shall show that this method also
yields more accurate results. In Wu, Chen and Chi\cite{Wu}(2005),
the authors also develop a differential theory for triangular
meshes. The gradient, Laplace-Beltrami operators are discussed.
Moreover, this method also works for boundary vertices and for
polyhedron meshes.

\end{document}